\documentclass[sowpacs,showkeys, groupedaddress,twocolumn,
floatfix, prb, aps]{revtex4-1}
\usepackage[english]{babel}
\usepackage{fontenc}
\usepackage{graphicx}
\usepackage{amsmath}
\usepackage{epstopdf}
\usepackage{amssymb}
\usepackage{amsbsy}
\usepackage{amscd}
\usepackage{float}
\usepackage{color}
\usepackage{tikz}
\usepackage{braket}
\usepackage{standalone}
\usepackage{bm}
\usepackage{siunitx}
\usepackage[normalem]{ulem} 
\usepackage{verbatim}
\usepackage{url}
\usetikzlibrary{positioning,calc}
\tikzset{>=latex}
\usepackage[verbose,hypertexnames=false,bookmarksopenlevel=1,filecolor=blue,
linkcolor=blue,citecolor=blue,urlcolor=blue,pdfstartview=FitH,bookmarksopen,bookmarksnumbered,
colorlinks,plainpages=false,linktocpage]{hyperref}
\begin{document}
\title{Hyperplane-Symmetric Static Einstein-Dirac Spacetime}
\author{John Schliemann and Tim Sonnleitner}
\affiliation{Institute for Theoretical Physics,
  University of Regensburg, Regensburg, Germany}
\date{\today}
\begin{abstract}
  We derive the general solution to the coupled Einstein and Dirac field
  equations in static and hyperplane-symmetric spacetime of arbitrary
  dimension including a cosmological constant of either sign. As a result,
  only a massful Dirac field couples via the Einstein equations to
  spacetime, and in the massless case the Dirac field is required to fulfill
  appropriate constraints in order to eliminate off-diagonal components of
  the energy-momentum tensor. We also give explicit expressions for curvature
  invariants including the Ricci scalar and the Kretschmann scalar, indicating
  physical singularities. Moreover, we reduce the general solution of the
  geodesic equation to quadratures.
\end{abstract}
\maketitle

\section{Introduction}
\label{Introduction}
The study of the Dirac field coupled to gravity, i.e. in curved spacetime
described by general relativity, has a long history, see, e.g.,
Refs.~\cite{Schrodinger32,Bargmann32,Brill57,Parker09,Collas18}.
As the coupled system of the Dirac equation and the Einstein field equations
is very intricate, mainly due to the nonlinear nature of the latter, explicit
solutions are usually available only under additional simplifying
assumptions.

An obvious option is to consider spacetimes being invariant under spatial
rotations.
\cite{Finster99,Ventrella03,Adanhounme12,Saha18,Blazquez-Salcedo19,Leith20,Bronnikov20,Kain23} A difficulty here is that a Dirac field carries a spin
$1/2$, so that the solutions of the Dirac equation necessarily
break rotational invariance. This problem can be overcome within appropriate
approximations, or by considering two distinct Dirac fields which couple
to a spin singlet.\cite{Finster99}
Spinor fields coupled to cylindrically spacetime were studied by
Saha.\cite{Saha20}.

In the present work we consider the coupled Einstein-Dirac system
in a hyperplane-symmetric spacetime of general dimension $D=1+d$. Here the
metric is invariant under the Euclidean group of a $(d-1)$-dimensional
hyperplane, i.e. under all translations, rotations, and reflections of that
plane. Such a spacetime was first studied by Taub\cite{Taub51} who found
the solution to the vacuum Einstein equations in dimension $D=1+3$.
Later Singh solved, in the same dimension, the Einstein equations coupled to
a massless Klein-Gordon field.\cite{Singh74} The latter study was generalized
by Vuille to the case of a finite cosmological constant.\cite{Vuille07}
Solutions of the Einstein equations in the presence of matter where later
also studied by Gomes.\cite{Gomes15}

This paper is organized as follows. In section \ref{Setup} we specify the
objects of our study with many technical details being deferred to
appendix \ref{TechnicalDetails}. The general solutions of the Einstein
field equations, depending on the sign of the cosmological constant and
the mass of the Dirac field, are developed in section
\ref{GeneralSolutionEinsteinFieldEquations}. The pertaining solutions
to the Dirac equation are described in section \ref{SolutionDiracEquation}.
Sections \ref{CurvatureInvariants} and \ref{Geodesics} contain discussions
of curvature invariants and geodesics, respectively for the metric tensors
derived before. We close with a summary and an outlook in section
\ref{SummaryOutlook}.
\section{Setup}
\label{Setup}
In curved spacetime, the Dirac equation for a spinor field $\psi(x)$ reads,
using standard notation,
\begin{equation}
  \left(i\tilde\gamma^{\mu}{\cal D}_{\mu}-m\right)\psi(x)=0\,,
  \label{diraceq1}
\end{equation}
with
\begin{equation}
  {\cal D}_{\mu}\psi=\left(\partial_{\mu}
  -\frac{1}{4}\omega_{\mu IJ}\gamma^I\gamma^J\right)\psi
  \label{covder}
\end{equation}
being the covariant derivative of the field, and
$\omega_{\mu IJ}$ is the usual spin connection. The
spacetime-dependent Dirac matrices $\tilde\gamma^{\mu}(x)=e^{\mu}_I(x)\gamma^I$
are given in terms of the $D$-bein and constant matrices $\gamma^I$
carrying internal indices and fulfilling the Dirac algebra,
\begin{equation}
  \left\{\gamma^I,\gamma^J\right\}=-2\eta^{IJ}{\mathbf 1}\,.
  \label{diracalg1}
\end{equation}
In general spacetime dimension $D=1+d$ we will not specify a representation
of this algebra, but we assume the usual (anti-)hermiticity properties,
\begin{equation}
  (\gamma^0)^+=\gamma^0
  \quad,\quad
  (\gamma^I)^+=-\gamma^I\,,\,I\in\{1,\dots,d\}\,.
  \label{diracalg2}
\end{equation}

To describe a static and hyperplane-symmetric spacetime we consider the metric
\begin{eqnarray}
  ds^2 & = & -e^{b(y)}(dx^0)^2+dy^2
  \nonumber\\
  & & \quad
  +e^{a(y)}\left((dx^2)^2+\cdots+(dx^d)^2\right)
  \label{metric1}
\end{eqnarray}
where the coordinate $x^1=y$ labels spatial hyperplanes of dimension
$d-1$. The metric is invariant under all rotations and translations of those
hyperplanes. An appropriate $D$-bein is chosen as
\begin{equation}
  e^{\mu}_I={\rm diag}\left(e^{-b(y)/2},1,
  e^{-a(y)/2},\dots,e^{-a(y)/2}\right)\,,
  \label{Dbein}
\end{equation}
and for all technical aspects not mentioned here we refer to appendix
\ref{TechnicalDetails}. As shown there, for the above geometrical data,
the Dirac equation (\ref{diraceq1}) for a
static spinor field $\psi(y)$ consistent with hyperplane symmetry
takes the form
\begin{equation}
  \psi^{\prime}=-\frac{1}{4}\left(b^{\prime}+(d-1)a^{\prime}\right)\psi
  +im\gamma^1\psi\,,
  \label{diraceq2}
\end{equation}
where the prime denotes differentiation with respect to $y$.
The Dirac-adjoint spinor $\bar\psi=\psi^+\gamma^0$ fulfills
\begin{equation}
  \bar\psi^{\prime}=-\frac{1}{4}\left(b^{\prime}+(d-1)a^{\prime}\right)\bar\psi
  -im\bar\psi\gamma^1
  \label{diraceq3}
\end{equation}
where we have used $(\gamma^1)^+=-\gamma^1$.
Using Eqs.~(\ref{diraceq2}), (\ref{diraceq3}) the nonvanishing components of
the energy-momentum tensor
\begin{eqnarray}
  T_{\mu\nu} & =- & \frac{i}{4}
  \Bigl(\bar\psi
  \left(\tilde\gamma_{\mu}{\cal D}_{\nu}
  +\tilde\gamma_{\nu}{\cal D}_{\mu}\right)\psi
  \nonumber\\
  & & \quad
  -\left({\cal D}_{\mu}\bar\psi\tilde\gamma_{\nu}
  +{\cal D}_{\nu}\bar\psi\tilde\gamma_{\mu}
  \right)\psi\Bigr)
  \label{ergmonten1}
\end{eqnarray}
can be identified as
\begin{equation}
  T_{11}=-m\bar\psi\psi
  \label{ergmonten2}
\end{equation}
and
\begin{equation}
  T_{0i}=-\frac{ie^{(b+a)/2}}{8}(b^{\prime}-a^{\prime})
  \bar\psi\gamma^0\gamma^1\gamma^I\psi\delta_{Ii}
  \quad,\quad
  i\geq 2\,.
  \label{ergmonten3}
\end{equation}

With the above findings the Einstein field equations (including a cosmological
constant $\Lambda$),
\begin{equation}
  R_{\mu\nu}-\frac{2}{D-2}\Lambda g_{\mu\nu}
  =\kappa\left(T_{\mu\nu}-\frac{T}{D-2}g_{\mu\nu}\right)\,,
  \label{einsteineq1}
\end{equation}
can be summarized as
\begin{eqnarray}
  R_{00} & = & \frac{e^b}{4}\left(2b^{\prime\prime}+b^{\prime 2}
  +(d-1)b^{\prime}a^{\prime}\right)\nonumber\\
  & = & -\frac{2\Lambda e^b}{d-1}-\kappa m\frac{e^b}{d-1}\bar\psi\psi\,,
  \label{einsteineq2}\\
  R_{11} & = & \frac{1}{4}\left(-2b^{\prime\prime}-b^{\prime 2}
  -(d-1)\left(2a^{\prime\prime}+a^{\prime 2}\right)\right)\nonumber\\
  & = & \frac{2\Lambda}{d-1}-\kappa m\frac{d-2}{d-1}\bar\psi\psi\,,
  \label{einsteineq3}\\
  R_{ii} & = & -\frac{e^a}{4}\left(2a^{\prime\prime}
  +b^{\prime}a^{\prime}+(d-1)a^{\prime 2}\right)\nonumber\\
  & = & \frac{2\Lambda e^a}{d-1}+\kappa m\frac{e^a}{d-1}\bar\psi\psi
  \quad,\quad i\geq 2\,.
  \label{einsteineq4}
\end{eqnarray}
In addition, since $R_{0i}=g_{0i}=0$ for $i\geq 2$,
the tensor component (\ref{ergmonten3}) is
required to vanish, which is the case if
\begin{equation}
  b(y)=a(y)+{\rm constant}\,,
  \label{einsteineq5}
\end{equation}
or
\begin{equation}
  \bar\psi\gamma^0\gamma^1\gamma^I\psi=0
  \quad,\quad
  I\geq 2\,.
  \label{einsteineq6}
\end{equation}
In the following section \ref{GeneralSolutionEinsteinFieldEquations}
we will see that the condition (\ref{einsteineq5})
is indeed fulfilled by the solution to the field equations
(\ref{einsteineq2})-(\ref{einsteineq4}) being consistent with the Dirac equation
(\ref{diraceq2}) for massive fields ($m>0$). Only in the massless case
$m=0$ the constraint (\ref{einsteineq6}) applies.
\section{General Solution to the Einstein Field Equations}
\label{GeneralSolutionEinsteinFieldEquations}
From Eqs.~(\ref{einsteineq2})-(\ref{einsteineq4}) it follows
\begin{eqnarray}
  & & \frac{1}{d-1}\left(e^{-b}R_{00}+R_{11}\right)+e^{-a}R_{ii}
  \nonumber\\
  & & \quad
  =-a^{\prime\prime}-\frac{d}{4}a^{\prime 2}
  =\frac{2\Lambda}{d-1}\,,
  \label{einsteinsol1}
\end{eqnarray}
and the latter equation, being of the Riccati type, can be solved by the
ansatz
\begin{equation}
  a^{\prime}(y)=\frac{4}{d}\frac{q^{\prime}(y)}{q(y)}
  \quad\Leftrightarrow\quad
  q(y)\propto e^{\frac{d}{4}a(y)}
  \label{einsteinsol2}
\end{equation}
leading to
\begin{equation}
  q^{\prime\prime}+\frac{d\Lambda}{2(d-1)}q=0\,.
  \label{einsteinsol3}
\end{equation}
Moreover, the Dirac equation (\ref{diraceq2}), (\ref{diraceq3}) implies
\begin{equation}
  \left(\bar\psi\psi\right)^{\prime}
  =-\frac{b^{\prime}+(d-1)a^{\prime}}{2}\bar\psi\psi
  \label{einsteinsol4}
\end{equation}
so that
\begin{equation}
  \left(\bar\psi\psi\right)(y)=\frac{C}{f(y)}
  \label{einsteinsol5}
\end{equation}
where $C$ is an integration constant $C$ and 
\begin{equation}
  f(y)=\exp\left(\frac{1}{2}\left(b(y)+(d-1)a(y)\right)\right)\,.
  \label{einsteinsol6}
\end{equation}
Using again the field equations one derives
\begin{eqnarray}
  4e^{-b}R_{00}+4R_{11} & = &
  (d-1)\left(b^{\prime}a^{\prime}-2a^{\prime\prime}-a^{\prime 2}\right)
  \nonumber\\
  & = & -\frac{4\kappa mC}{f}
  \label{einsteinsol7}
\end{eqnarray}
leading to an equation for the function (\ref{einsteinsol6})\,,
\begin{equation}
  f^{\prime}
  -\left(\frac{d}{4}a^{\prime}-\frac{2\Lambda}{d-1}\frac{1}{a^{\prime}}\right)f
  =-\frac{2\kappa mC}{d-1}\frac{1}{a^{\prime}}\,.
  \label{einsteinsol8}
\end{equation}
\subsection{Negative Cosmological Constant $\Lambda<0$}
\label{Lambda<0}
Let us first discuss the case of a finite negative cosmological constant
$\Lambda\neq 0$. Here one readily finds
from Eqs.~(\ref{einsteinsol2}), (\ref{einsteinsol3})
\begin{equation}
  e^{a(y)}=A\left(\cosh(Y(y))\right)^{\frac{4}{d}}
  \label{einsteinsol9}
\end{equation}
with
\begin{equation}
  Y(y)=\sqrt{\frac{d|\Lambda|}{2(d-1)}}y+\eta
  \label{einsteinsol10}
\end{equation}
and two integration constants $\eta$ and $A\geq 0$.

The homogeneous ($m=0$) part of Eq.~(\ref{einsteinsol8}),
\begin{equation}
  f^{\prime}_0
  -\left(\frac{d}{4}a^{\prime}+\frac{2|\Lambda|}{d-1}\frac{1}{a^{\prime}}\right)
  f_0=0\,,
  \label{einsteinsol11}
\end{equation}
is solved by 
\begin{eqnarray}
  f_0(y) & = & \exp\left(\int^yd\bar y
  \left(\frac{d}{4}a^{\prime}(\bar y)
  +\frac{2|\Lambda|}{d-1}\frac{1}{a^{\prime}(\bar y)}\right)
  \right)\nonumber\\
  & = & E\left|\cosh(Y(y))\sinh(Y(y))\right|
  \label{einsteinsol12}
\end{eqnarray}
with another integration constant $E$. Note the modulus occurring in the
above result which will become important in what follows.
\subsubsection{Massive Dirac Field: $m>0$}
\label{Lambda<0m>0}
Turning first to the case of a massful Dirac field, $m>0$,
a special solution to Eq.~(\ref{einsteinsol8}) can be given as
\begin{equation}
  f_1(y)=-\frac{2\kappa mC}{d-1}f_0(y)
  \int^yd\bar y\frac{1}{a^{\prime}(\bar y)f_0(\bar y)}
  \label{einsteinsol13}
\end{equation}
with
\begin{eqnarray}
  & &
  \int^yd\bar y\frac{1}{a^{\prime}(\bar y)f_0(\bar y)}
  \nonumber\\
  & & \quad
  =-\frac{1}{E}\frac{d-1}{2|\Lambda|}\left|\coth(Y(y))\right|+{\rm constant}\,,
  \label{einsteinsol14}
\end{eqnarray}
so that the general solution of Eq.~(\ref{einsteinsol8}) is
\begin{eqnarray}
  f(y) & = & \left|\cosh(Y(y))\sinh(Y(y))\right|
  \nonumber\\
  & & \cdot
  \left(E+\frac{\kappa mC}{|\Lambda|}\left|\coth(Y(y))\right|\right)\,.
  \label{einsteinsol15}
\end{eqnarray}
Note that the above function is by construction positive which puts
restrictions on $E,\kappa mC$.
Thus, solving for $b(y)$,
\begin{eqnarray}
   e^{b(y)} & = & \frac{1}{A^{d-1}}\left(\cosh(Y(y))\right)^{\frac{4}{d}}
  \nonumber\\
  & & \cdot
  \left(E\left|\tanh(Y(y))\right|+\frac{\kappa mC}{|\Lambda|}\right)^2\,.
  \label{einsteinsol16}
\end{eqnarray}
The derivatives of $b(y)$ are easily computed as
\begin{eqnarray}
  b^{\prime} & = & \left(\frac{4}{d}\frac{s}{c}
  +\frac{2\nu E\frac{1}{c^2}}{E\left|\frac{s}{c}\right|
    +\frac{\kappa mC}{|\Lambda|}}\right)
  \sqrt{\frac{d|\Lambda|}{2(d-1)}}\,,
  \label{einsteinsol17}\\
  b^{\prime\prime} & = & \Biggl(\frac{4}{d}\frac{1}{c^2}
  +\frac{d\nu}{dY}\frac{2E\frac{1}{c^2}}{E\left|\frac{s}{c}\right|
    +\frac{\kappa mC}{|\Lambda|}}
  -\frac{4E\frac{1}{c^2}\left|\frac{s}{c}\right|}{E\left|\frac{s}{c}\right|
    +\frac{\kappa mC}{|\Lambda|}}
  \nonumber\\
  & & \quad
  -\frac{2E^2\frac{1}{c^4}}
  {\left(E\left|\frac{s}{c}\right|+\frac{\kappa mC}{|\Lambda|}\right)^2}
  \Biggr)\frac{d|\Lambda|}{2(d-1)}\,,
  \label{einsteinsol18}
\end{eqnarray}
where we have defined $c=\cosh(Y(y))$, $s=\sinh(Y(y))$, and
$\nu(y)={\rm sign}(Y(y))$. Thus, the derivative of $\nu$ entering
Eq.~(\ref{einsteinsol18}) is a $\delta$-peak located at $y=y_0$  where
$Y(y_0)=0$.

Inserting now the above results into the field equation (\ref{einsteineq2})
leads to 
\begin{eqnarray}
  & & \frac{1}{4}\left(2b^{\prime\prime}+b^{\prime 2}
  +(d-1)b^{\prime}a^{\prime}\right)
  \nonumber\\
  & & \quad
  =\left(\frac{d\nu}{dY}
  \frac{\frac{d}{2}E\frac{1}{c^2}}{E\left|\frac{s}{c}\right|
    +\frac{\kappa mC}{|\Lambda|}}+2
  -\frac{\frac{\kappa mC}{|\Lambda|}\frac{1}{c^2}}{E\left|\frac{s}{c}\right|
    +\frac{\kappa mC}{|\Lambda|}}\right)\frac{|\Lambda|}{d-1}
  \nonumber\\
  & & \quad
  =\frac{2|\Lambda|}{d-1}-\frac{\kappa mC}{d-1}
    \frac{1}{c^2\left(E\left|\frac{s}{c}\right|
      +\frac{\kappa mC}{|\Lambda|}\right)}\,,
    \label{einsteinsol19}
\end{eqnarray}
which is fulfilled if and only if the term proportional to the
$\delta$-peak $d\nu/dY$ vanishes. Therefore the integration constant
$E$ has to be chosen to be zero, and consistent conclusions follow from the
field equations (\ref{einsteineq3}), (\ref{einsteineq4}). Thus, we have,
along with the expression (\ref{einsteinsol9}),
\begin{eqnarray}
  e^{b(y)} & = & B\left(\cosh(Y(y))\right)^{\frac{4}{d}}\,.
  \label{einsteinsol20}\\
  \left(\bar\psi\psi\right)(y) & = & \frac{\frac{|\Lambda|}{\kappa m}}
       {\cosh^2(Y(y)))}
  \label{einsteinsol21}
\end{eqnarray}
with $B=(\kappa mC/|\Lambda|)^2A^{1-d}$. Moreover, Eqs.~(\ref{einsteinsol9})
and (\ref{einsteinsol20}) satisfy the condition (\ref{einsteineq5}). Hence,
we have found the general solution to the Einstein field  equations coupled to
a Dirac field in the presence of a negative cosmological constant.
\subsubsection{Massless Dirac Field: $m=0$}
\label{Lambda<0m=0}
For zero mass $m=0$ Eq.~(\ref{einsteinsol16}) turns into
\begin{equation}
  e^{b(y)}=B\left(\cosh(Y(y))\right)^{\frac{4}{d}}\tanh^2(Y(y))
  \label{einsteinsol21a}
\end{equation}
with $B=EA^{1-d}$. Here no $\delta$-singularities occur in the derivatives of
$b(y)$, and Eq.~(\ref{einsteinsol21}) becomes
\begin{equation}
  \left(\bar\psi\psi\right)(y)
  =\frac{C}{\left|\cosh(Y(y))\sinh(Y(y))\right|}
  \label{einsteinsol21b}
\end{equation}
with $C/E\mapsto C$ being another free integration constant.
This is a natural finding since for $m=0$ the Dirac field does not enter the
Einstein field equations 
The expressions (\ref{einsteinsol9}) and (\ref{einsteinsol21a})
solve the field equations (\ref{einsteineq2})-(\ref{einsteineq4}) for $m=0$,
but do not obey the condition (\ref{einsteineq5}). Hence, to provide a full
solution of the field equations, the Dirac field must fulfill the
constraint (\ref{einsteineq6}).
Note that, similarly to Eq.~(\ref{einsteinsol4}),
the Dirac equation (\ref{diraceq2}), (\ref{diraceq3}) leads to
\begin{equation}
  \left(\bar\psi\gamma^0\gamma^1\gamma^I\psi\right)^{\prime}
  =-\frac{b^{\prime}+(d-1)a^{\prime}}{2}\bar\psi\gamma^0\gamma^1\gamma^I\psi\,.
  \label{einsteinsol21c}
\end{equation}
Moreover, the l.h.s. of Eq.~(\ref{einsteineq6}) is by the general
properties (\ref{diracalg1}), (\ref{diracalg2}) restricted to be purely
imaginary. Thus, this equation provides $(d-1)$ real conditions on the
complex Dirac field $\psi(y)$.
We will discuss the constraint (\ref{einsteineq6}) further
in section \ref{SolutionDiracEquationm=0}.
\subsection{Positive Cosmological Constant $\Lambda>0$}
\label{Lambda>0}
For finite positive cosmological constant $\Lambda>0$, Eq.~(\ref{einsteinsol3})
is solved by trigonometric functions, instead of hyperbolic ones as for
$\Lambda<0$, and the general solution to Eq.~(\ref{einsteinsol1}) is
\begin{equation}
  e^{a(y)}=A\left|\cos(Y(y))\right|^{\frac{4}{d}}
  \label{einsteinsol22}
\end{equation}
with two integration constants as in Eqs.~(\ref{einsteinsol9}),
(\ref{einsteinsol10}). Indeed, the following analysis proceeds fairly
analogous to the previous case $\Lambda<0$ with hyperbolic expressions
to be replaced with appropriate trigonometric ones.
\subsubsection{Massive Dirac Field: $m>0$}
\label{Lambda>0m>0}
The  general solution of Eq.~(\ref{einsteinsol8}) reads for positive
$\lambda>0$
\begin{eqnarray}
  f(y) & = & \left|\cos(Y(y))\sin(Y(y))\right|
  \nonumber\\
  & & \cdot
  \left(E-\frac{\kappa mC}{\Lambda}\left|\cot(Y(y))\right|\right)\,,
  \label{einsteinsol23}
\end{eqnarray}
so that
\begin{eqnarray}
   e^{b(y)} & = & \frac{1}{A^{d-1}}\left|\cos(Y(y))\right|^{\frac{4}{d}}
  \nonumber\\
  & & \cdot
  \left(E\left|\tan(Y(y))\right|-\frac{\kappa mC}{\Lambda}\right)^2\,.
  \label{einsteinsol24}
\end{eqnarray}
Similarly as in Eq.~(\ref{einsteinsol18}), the second derivative
$b^{\prime\prime}(y)$ contains singularities which occur here at
$y=y_n$ with $Y(y_n)=(\pi/2)n$, $n\in\mathbb{Z}$. Again consistency with
the field equations (\ref{einsteineq2})-(\ref{einsteineq4}) requires
these singularities to vanish so that the integration constant $E$ is zero.
In summary, we have
\begin{eqnarray}
  e^{b(y)} & = & B\left|\cos(Y(y))\right|^{\frac{4}{d}}\,.
  \label{einsteinsol25}\\
  \left(\bar\psi\psi\right)(y) & = & \frac{-\frac{\Lambda}{\kappa m}}
       {\cos^2(Y(y)))}
  \label{einsteinsol26}
\end{eqnarray}
with again $B=(\kappa mC/\Lambda)^2A^{1-d}$.
Finally, the expressions (\ref{einsteinsol22})
and (\ref{einsteinsol25}) fulfill the condition (\ref{einsteineq5}). 
\subsubsection{Massless Dirac Field: $m=0$}
\label{Lambda>0m=0}
Analogously as in section \ref{Lambda<0m=0} it follows
for a massless Dirac field,
instead of the expressions (\ref{einsteinsol25}), (\ref{einsteinsol26}),
\begin{eqnarray}
  e^{b(y)} & = & B\left|\cos(Y(y))\right|^{\frac{4}{d}}\tan^2(Y(y))\,.
  \label{einsteinsol26a}\\
  \left(\bar\psi\psi\right)(y) & = &
  \frac{C}{\left|\cos(Y(y))\sin(Y(y))\right|}
  \label{einsteinsol26b}
\end{eqnarray}
with two integration constants $B$, $C$. Again, the results
(\ref{einsteinsol22}), (\ref{einsteinsol26a}) fail to fullfill the
condition (\ref{einsteineq5}), the constraint (\ref{einsteineq6})
applies.
\subsection{Zero Cosmological Constant $\Lambda=0$}
\label{Lambda=0}
For a vanishing cosmological constant one finds, proceeding as in the previous
sections,
\begin{equation}
  e^{a(y)}=\tilde A\left|y+\eta\right|^{\frac{4}{d}}\,,
  \label{einsteinsol27}
\end{equation}
with some integration constant $\tilde A$.
Note that the marginal case $\Lambda=0$ is unstable in the sense that
any small but finite cosmological constant (of either sign) qualitatively
changes the solution.
\subsubsection{Massive Dirac Field: $m>0$}
\label{Lambda=0m>0}
For a massful Dirac field the remaining quantities entering the Einstein
field  equations (\ref{einsteineq2})-(\ref{einsteineq4}) are
found as, analogously as in the previous sections,
\begin{eqnarray}
  e^{b(y)} & = & \tilde B\left|y+\eta\right|^{\frac{4}{d}}\,,
  \label{einsteinsol28}\\
  \left(\bar\psi\psi\right)(y) & = & 
  -2\frac{d-1}{d}\frac{\frac{1}{\kappa m}}
       {(y+\eta)^2}
  \label{einsteinsol29}
\end{eqnarray}
with two more dimensionful integration constants $\tilde B\geq 0$ and
$\eta$. Again the above expressions satisfy the condition (\ref{einsteineq5}).
\subsubsection{Massless Dirac Field: $m=0$}
\label{Lambda=0m=0}
For a Dirac field of vanishing mass $m=0$, Eqs.~(\ref{einsteinsol28}),
(\ref{einsteinsol29}) turn into
\begin{eqnarray}
  e^{b(y)} & = & \tilde B\left|y+\eta\right|^{\frac{4}{d}}
  \left(y+\eta\right)^{-2}\,,
  \label{einsteinsol30}\\
  \left(\bar\psi\psi\right)(y) & = & 
  \frac{C}{|y+\eta|}\,.
  \label{einsteinsol31}
\end{eqnarray}
Again, the results
(\ref{einsteinsol27}), (\ref{einsteinsol30}) do not obey the
condition (\ref{einsteineq5}), so that the Dirac field must fulfill
the constraint (\ref{einsteineq6}).
\section{Solution to the Dirac Equation}
\label{SolutionDiracEquation}
In the general solution of the Einstein field equations, the spinor field
$\psi(y)$ enters only in terms of the quantity $\bar\psi\psi$. To investigate
further the solutions to the Dirac equation we specify the spacetime dimension
to be $D=1+3$ and use the standard representation of the Dirac algebra,
\begin{equation}
\gamma^0=\left(
  \begin{array}{cc}
    {\mathbf 1} & 0\\
    0 & -{\mathbf 1}
  \end{array}
  \right)
  \quad,\quad
  \gamma^I=\left(
  \begin{array}{cc}
    0 & \sigma^I \\
    -\sigma^I & 0
  \end{array}
  \right)
  \label{diracsol1}
\end{equation}
with $I\in\{1,2,3\}$ and the usual Pauli matrices
\begin{equation}
\sigma^1=\left(
  \begin{array}{cc}
    0 & 1\\
    1 & 0
  \end{array}
  \right)\,,\,
\sigma^2=\left(
  \begin{array}{cc}
    0 & -i\\
    i & 0
  \end{array}
  \right)\,,\,
  \sigma^3=\left(
  \begin{array}{cc}
    1 & 0\\
    0 & -1
  \end{array}
  \right)\,.
  \nonumber\\
  \label{diracsol2}
\end{equation}
\subsection{Massive Dirac Field: $m>0$}
\label{SolutionDiracEquationm>0}
Employing the eigenspinors of the Hermitian matrix $i\gamma^1$ one
constructs four linearly independent solutions to the Dirac equation
(\ref{diraceq2}) for a massive field,
\begin{eqnarray}
	\lambda_{\pm}(y) & = & \frac{1}{\sqrt{g(y)}}
	\left(
	\begin{array}{c}
		1 \\
		0 \\
		0 \\
		\pm i
	\end{array}
	\right)e^{\mp m y}\,,
	\label{diracsol3}\\
	\kappa_{\pm}(y) & = &  \frac{1}{\sqrt{g(y)}}
	\left(
	\begin{array}{c}
		0 \\
		1 \\
		\pm i \\
		0
	\end{array}
	\right)e^{\mp m y} \, .
	\label{diracsol4}
\end{eqnarray}
with
\begin{equation}
  g(y)=\left\{
  \begin{array}{cc}
    \cosh^2(Y(y)) & \Lambda<0 \\
    \cos^2(Y(y)) & \Lambda>0 \\
    (y+\eta)^2 & \Lambda=0
  \end{array}
  \right.\,.
  \label{diracsol5}
\end{equation}
Thus, the general solution is
\begin{equation}
	\psi=C_+\lambda_++C_-\lambda_-+D_+\kappa_++D_-\kappa_-
	\label{diracsol6}
\end{equation}
where the integration constants $C_{\pm}$, $D_{\pm}$ fulfill according to
Eqs.~(\ref{einsteinsol21}), ~(\ref{einsteinsol26}), ~(\ref{einsteinsol29})
\begin{equation}
  2{\rm Re}\left\{C^*_+C_-+D^*_+D_-\right\}
  =\left\{
  \begin{array}{cc}
    -\frac{\Lambda}{\kappa m} & \Lambda\neq 0 \\
    -\frac{2(d-1)}{d}\frac{1}{\kappa m} & \Lambda=0
  \end{array}
    \right.\,.
  \label{diracsol7}
\end{equation}
The above condition must be satisfied for  consistency of the solutions to
the Dirac equation with the Einstein field equations.
\subsection{Massless Dirac Field: $m=0$}
\label{SolutionDiracEquationm=0}
For a massless Dirac field, the basis solutions (\ref{diracsol3}),
(\ref{diracsol4}) are to be modified as
\begin{eqnarray}
	\lambda_{\pm}(y) & = & \frac{1}{\sqrt{\bar g(y)}}
	\left(
	\begin{array}{c}
		1 \\
		0 \\
		0 \\
		\pm i
	\end{array}
	\right)\,,
	\label{diracsol8}\\
	\kappa_{\pm}(y) & = &  \frac{1}{\sqrt{\bar g(y)}}
	\left(
	\begin{array}{c}
		0 \\
		1 \\
		\pm i \\
		0
	\end{array}
	\right)\, .
	\label{diracsol9}
\end{eqnarray}
with
\begin{equation}
  \bar g(y)=\left\{
  \begin{array}{cc}
    |\cosh(Y(y))\sinh(Y(y))| & \Lambda<0 \\
    |\cos(Y(y))\sin(Y(y))| & \Lambda>0 \\
    |y+\eta| & \Lambda=0
  \end{array}
  \right.\,.
  \label{diracsol10}
\end{equation}
The general solution is again of the form (\ref{diracsol6}) with the integration
constants fulfilling, according to
Eqs.~(\ref{einsteinsol21b}), (\ref{einsteinsol26b}), (\ref{einsteinsol31}),
\begin{equation}
  2{\rm Re}\left\{C^*_+C_-+D^*_+D_-\right\}=C\,.
  \label{diracsol11}
\end{equation}

Moreover, as seen in sections \ref{Lambda<0m=0}, \ref{Lambda>0m=0},
and \ref{Lambda=0m=0}, the massless Dirac field has to obey the
constraint (\ref{einsteineq6}).
Formulating the $4$-spinor as $\psi=(\chi_1,\chi_2)^T$, this condition
reads in terms of the $2$-spinors $\chi_{1/2}$
\begin{equation}
  \chi^+_1\sigma^I\chi_1+\chi^+_2\sigma^I\chi_2=0
  \quad,\quad
  I\in\{2,3\}\,.
  \label{diracsol12}
\end{equation}
The latter equations imply, as detailed in appendix \ref{TheConstraint},
for $I=2$
\begin{equation}
	{\rm Im}\left\{C^*_+D_--D^*_+C_-\right\}=0\,,
	\label{diracsol13}
\end{equation}
while for $I=3$ it follows
\begin{equation}
	{\rm Re}\left\{C^*_+C_--D^*_+D_-\right\}=0\,.
	\label{diracsol14}
\end{equation}
The above equations are two real conditions on the four complex
quantities $C_{\pm}$, $D_{\pm}$. Thus, the resulting manifold of solutions
is described by six real parameters.
\section{Curvature Invariants}
\label{CurvatureInvariants}
We now discuss curvature invariants of the spacetimes derived in section
\ref{GeneralSolutionEinsteinFieldEquations}.

From the field equations (\ref{einsteineq1}) and Eq.~(\ref{ergmonten1})
one easily derives a general expression for the Ricci scalar,
\begin{eqnarray}
  R & = & \frac{2D}{D-2}\Lambda-\frac{2\kappa}{D-2}T
  \nonumber\\  & = & 
  \frac{2}{d-1}\left((d+1)\Lambda+\kappa m\bar\psi\psi\right)\,.
  \label{curvinv1}
\end{eqnarray}
Likewise, for the square of the Ricci tensor it follows
\begin{eqnarray}
  R^{\mu\nu}R_{\mu\nu} & = &
  \frac{4D}{(D-2)^2}\Lambda^2
  +\frac{4\Lambda\kappa}{D-2}\left(T-\frac{D}{D-2}T\right)
  \nonumber\\
  & & \quad
  +\kappa^2\left(T^{\mu\nu}T_{\mu\nu}-\frac{2}{D-2}T^2+\frac{D}{(D-2)^2}T^2\right)
  \nonumber\\
  & = & 
  \frac{1}{(d-1)^2}
  \Bigl(
  4(d+1)\Lambda^2+8\Lambda\kappa m\bar\psi\psi
  \nonumber\\
  & & \quad
  +\kappa^2m^2\left(d^2-3d+4\right)\left(\bar\psi\psi\right)^2
  \Bigr)\,.
  \label{curvinv2}
\end{eqnarray}
Here we have used $T^{\mu\nu}T_{\mu\nu}=T^2$ which holds true if one of the
conditions (\ref{einsteineq5}). (\ref{einsteineq6}) is fulfilled so that the
only nonzero component of the energy-momentum tensor is given by
Eq.~(\ref{ergmonten2}).

Thus, the potential sigularities of both quantities (\ref{curvinv1}),
(\ref{curvinv2}) are the singularities of the bilinear form
$\bar\psi\psi$. For the Ricci scalar one finds
via Eqs.~(\ref{einsteinsol21}), ~(\ref{einsteinsol26}), ~(\ref{einsteinsol29})
for the massful case ($m>0$)
\begin{equation}
  R=\left\{
  \begin{array}{cc}
    \frac{2\Lambda}{d-1}
    \left(d+1-\frac{1}{g(y)}\right) & \Lambda\neq 0 \\
    -\frac{4}{d}\frac{1}{g(y)} & \Lambda=0
  \end{array}
  \right.
  \label{curvinv3}
\end{equation}
with $g(y)$ as in Eq.~(\ref{diracsol5}), whereas for zero mass $m=0$ it follows
directly from Eq.~(\ref{curvinv1})
\begin{equation}
  R=2\frac{d+1}{d-1}\Lambda\,.
  \label{curvinv4}
\end{equation}
Thus, in the case of a massless Dirac field the Ricci scalar is trivially
constant, and the same holds true for the square of the Ricci tensor
as easily seen by putting $m=0$  in Eq.~(\ref{curvinv2}).
To further analyze possible curvature singularities in the massless  
case we turn to the Kretschmann scalar (\ref{curv6}) with the result
\begin{equation}
	K=\left\{
	\begin{array}{cc}
		\frac{4\Lambda^2}{d \left(d-1\right)} \left(\frac{\left(d-1\right) \left(d-2\right)}{\bar{g}^2} + 2\right) & \Lambda\neq 0 \\
		\frac{16 \left(d-2\right) \left(d-1\right)^2}{d^3}\frac{1}{\bar{g}^2} & \Lambda=0
	\end{array}
	\right. \,,
	\label{curvinv5}
\end{equation}
where $\bar g(y)$ is given by Eq.~(\ref{diracsol10}),
and the above invariant obviously diverges at the zeros of this function.

In the massful case the Kretschmann scalar (\ref{curv6}) simplifies
due to the condition (\ref{einsteineq5}) so that
\begin{equation}
  K=d\left(\left(b^{\prime\prime}+\frac{b^{\prime 2}}{2}\right)^2
  +\frac{d-1}{2}\frac{b^{\prime 4}}{4}\right)\,.
  \label{curvinv6}
\end{equation}
When evaluated further this quantity shows singularities at the same positions
as the Ricci scalar (\ref{curvinv3}).

As the quantities (\ref{curvinv3}), (\ref{curvinv5}) are invariant under
arbitrary coordinate transformations, their singularities are of physical
nature and not coordinate singularities.
\section{Geodesics}
\label{Geodesics}
We now analyze the solutions of the geodesic equation
\begin{equation}
  u^{\nu}\nabla_{\nu}u^{\mu}
  =\frac{du^{\mu}}{d\tau}+\Gamma^{\mu}_{\nu\kappa}u^{\nu}u^{\kappa}=0\,.
  \label{geo1}
\end{equation}
for the hyperplane-symmetric spacetimes studied in section
\ref{GeneralSolutionEinsteinFieldEquations}.
Here the velocity $u^{\mu}=dx^{\mu}/d\tau$ is parameterized by an affine
parameter $\tau$ so that
\begin{equation}
  u^{\mu}u_{\mu}=-\varepsilon
  =\left\{
  \begin{array}{cc}
    -1 & {\rm timelike} \\
    0 & {\rm lightlike} \\
    1 & {\rm spacelike} 
  \end{array}
  \right.
  \label{geo2}
\end{equation}
where the parameter $\varepsilon\in\{-1,0,1\}$ distinguishes the type of
geodesic as stated above.

The geodesic equation (\ref{geo1}) reads componentwise
\begin{eqnarray}
  \frac{du^0}{d\tau} & = & -b^{\prime}u^0u^1\,,
  \label{geo3}\\
  \frac{du^1}{d\tau} & = &
  -e^b\frac{b^{\prime}}{2}\left(u^0\right)^2
  +e^a\frac{a^{\prime}}{2}\sum_{i=2}^d\left(u^i\right)^2\,,
  \label{geo4}\\
  \frac{du^i}{d\tau} & = & -a^{\prime}u^1u^i
  \quad,\quad
  i\geq 2\,.
  \label{geo5}
\end{eqnarray}
The hyperplane-symmetric metric admits a timelike Killing vector,
\begin{equation}
  ^{(0)}\xi_{\mu}=\left(-e^b,0,\dots,0\right)\,,
  \label{geo6}
\end{equation}
and $d-1$ spacelike Killing vectors in the hyperplane,
\begin{equation}
  ^{(i)}\xi_{\mu}=\left(0,0,0,\dots,e^a\dots,0\right)
  \quad,\quad
  i\geq 2\,,
  \label{geo7}
\end{equation}
all satisfying the Killing equation
$\nabla_{\mu}\xi_{\nu}+\nabla_{\nu}\xi_{\mu}=0$. The corresponding conserved
quantities are
\begin{eqnarray}
  k^{(0)} & = & {^{(0)}\xi_{\mu}}u^{\mu}=-u^0e^b\,,
  \label{geo8}\\
  k^{(i)} & = & {^{(i)}\xi_{\mu}}u^{\mu}=u^ie^a
  \quad,\quad
  i\geq 2\,,
  \label{geo9}
\end{eqnarray}
which can of course also be derived directly from Eqs.~(\ref{geo3}),
(\ref{geo5}). Note that the Killing vectors (\ref{geo7}) refer to spatial
translations, and there are $(d-1)(d-2)/2$ more Killing vectors describing
rotations of the hyperplane, which, however, do not lead to new conclusions.

Eqs.~(\ref{geo4}), (\ref{geo2}) can now be reformulated as
\begin{eqnarray}
  \frac{du^1}{d\tau} & = &
  -e^{-b}\frac{b^{\prime}}{2}\left(k^{(0)}\right)^2
  +e^{-a}\frac{a^{\prime}}{2}\sum_{i=2}^d\left(k^{(i)}\right)^2\,,
  \label{geo10}\\
  \left(u^1\right)^2 & = &
  -\varepsilon
  +e^{-b}\left(k^{(0)}\right)^2
  -e^{-a}\sum_{i=2}^d\left(k^{(i)}\right)^2\,,
  \label{geo11}
\end{eqnarray}
where the latter equation is an integrated version of the former with an
appropriate integration constant $(-\varepsilon)$. Thus, we have found all
$D=1+d$ integrals of motion of the geodesic equation (\ref{geo1}).

The coordinate $y(\tau)$ can now formally be expressed as
\begin{eqnarray}
  & & \int_{y(\tau_0)}^{y(\tau)}
  \frac{d\bar y}{\sqrt{-\varepsilon
  +e^{-b(\bar y)}\left(k^{(0)}\right)^2
  -e^{-a(\bar y)}\sum_{i=2}^d\left(k^{(i)}\right)^2}}
  \nonumber\\
  & & \qquad=\pm(\tau-\tau_0)\,.
  \label{geo12}
\end{eqnarray}
In the case of a massfull Dirac field the condition (\ref{einsteineq5}) is
satisfied, and the above expression simplifies to
\begin{equation}
  \int_{y(\tau_0)}^{y(\tau)}
  \frac{d\bar y}{\sqrt{-\varepsilon
  +M(g(\bar y))^{-2/d}}}
  =\pm(\tau-\tau_0)
  \label{geo13}
\end{equation}
with some constant $M$ and $g(y)$ given by Eq.~(\ref{diracsol5}).
\section{Summary and Outlook}
\label{SummaryOutlook}
We have analyzed the coupled Einstein and Dirac field
equations in static and hyperplane-symmetric spacetime of arbitrary
dimension in the presence of a cosmological constant.
This geometry allows for complete explicit analytic expressions
for the metric tensor and the Dirac field.
Regarding the cosmological constant, three cases are to be distinguished
(negative/positive/zero). Moreover, only a massful Dirac field couples via the
Einstein equations to spacetime, and in the massless case the Dirac field is
required to fulfill the constraints (\ref{einsteineq6}) 
in order to ensure that off-diagonal components of the energy-momentum tensor
vanish.

The singularities of the metric tensors are of physical nature, as
indicated by curvature invariants such as the Ricci scalar and the
Kretschmann scalar.
Finally, we have reduced, making use of geometric symmetries,
the general solution of the geodesic equation to quadratures.

Hyperplane-symmetric spacetime coupled to the Klein-Gordon field has been
studied earlier\cite{Singh74,Vuille07}, and the present work adds the case
of the Dirac field.
It would be interesting to see whether fields describing objects of higher spin,
such as the Proca and Rarita-Schwinger field, allow for similarly explicit
analytic results as found here.

\appendix
\section{Technical Details}
\label{TechnicalDetails}
\subsection{Christoffel Symbols and Curvature Tensor}
\label{ChristoffelTensor}

The nonvanishing Christoffel symbols according  to the line element
(\ref{metric1}) are
\begin{eqnarray}
  & &
  \Gamma^0_{01}=\Gamma^0_{10}=\frac{b^{\prime}}{2}
  \quad,\quad
  \Gamma^1_{00}=e^b\frac{b^{\prime}}{2}\,,
  \label{christoffel1}\\
  & & 
  \Gamma^1_{ii}=-e^a\frac{a^{\prime}}{2}
  \quad,\quad
  \Gamma^i_{1i}=\Gamma^i_{i1}=\frac{a^{\prime}}{2}
  \,\,,\,\,
  i\geq  2\,,
  \label{christoffel2}
\end{eqnarray}
so that the nonzero and independent components of the curvature tensor
\begin{equation}
  {R_{\mu\nu\kappa}}^{\lambda}=\partial_{\nu}\Gamma^{\lambda}_{\mu\kappa}
  -\partial_{\mu}\Gamma^{\lambda}_{\nu\kappa}
  +\Gamma^{\rho}_{\mu\kappa}\Gamma^{\lambda}_{\nu\rho}
  -\Gamma^{\rho}_{\nu\kappa}\Gamma^{\lambda}_{\mu\rho}
\label{curv1}
\end{equation}
can be summarized as
\begin{eqnarray}
  R_{0101} & = & e^b\left(\frac{b^{\prime\prime}}{2}+\frac{b^{\prime 2}}{4}\right)
  \label{curv2}\\
  R_{0i0i} & = & e^{(b+a)}\frac{b^{\prime}a^{\prime}}{4}
  \quad,\quad
  i\geq  2\,,
  \label{curv3}\\
  R_{1i1i} & = & -e^a\left(\frac{a^{\prime\prime}}{2}+\frac{a^{\prime 2}}{4}\right)
  \quad,\quad
  i\geq  2\,,
  \label{curv4}\\
  R_{ijij} & = & -e^{2a}\frac{a^{\prime 2}}{4}
  \quad,\quad
  i,j\geq  2\,,\,i\neq j\,.
  \label{curv5}
\end{eqnarray}
These expressions lead to the Ricci tensor components given on the r.h.s.
of Eqs.~(\ref{einsteineq2})-(\ref{einsteineq4}). Moreover, for the Kretschmann
scalar one finds
\begin{eqnarray}
  & & K=R^{\mu\nu\kappa\lambda}R_{\mu\nu\kappa\lambda}
  \nonumber\\
  & & 
  =\left(b^{\prime\prime}+\frac{b^{\prime 2}}{2}\right)^2
  +(d-1)\left(\left(\frac{b^{\prime}a^{\prime}}{2}\right)^2
  + \left(a^{\prime\prime}+\frac{a^{\prime 2}}{2}\right)^2\right)
  \nonumber\\
  & & \quad\quad 
  +\frac{(d-1)(d-2)}{2}\frac{a^{\prime  4}}{4}\,.
  \label{curv6}
\end{eqnarray}
\subsection{Spin Connection and Dirac Equation}
\label{SpinConnectionDiracEquation}

The nonvanishing usual covariant derivatives
\begin{equation}
  \nabla_{\mu}e^{\nu}_I=\partial_{\mu}e^{\nu}_I+\Gamma^{\nu}_{\mu\kappa}e^{\kappa}_I
  \label{spincon1}
\end{equation}
of the inverse $D$-bein (\ref{Dbein}) are
\begin{eqnarray}
  \nabla_0e^0_I & = & \frac{b^{\prime}}{2}\delta_{I1}\,,
  \label{spincon2}\\
  \nabla_0e^1_I & = & e^{b/2}\frac{b^{\prime}}{2}\delta_{I0}\,,
  \label{spincon3}\\
  \nabla_ie^1_I & = & -e^{a/2}\frac{a^{\prime}}{2}\delta_{iI}
  \quad,\quad i\geq 2\,,
  \label{spincon4}\\
  \nabla_ie^i_I & = & \frac{a^{\prime}}{2}\delta_{1I}
  \quad,\quad i\geq 2\,,
  \label{spincon5}
\end{eqnarray}
Using this data, the nonzero components of the spin connection
\begin{equation}
  {{\omega_{\mu}}^I}_J=e^I_{\nu}\nabla_{\mu}e^{\nu}_J
  \label{spincon6}
\end{equation}
are obtained as
\begin{eqnarray}
  \omega_{001}=-\omega_{010} & = & -e^{b/2}\frac{b^{\prime}}{2}\,,
  \label{spincon8}\\
  \omega_{i1I}=-\omega_{iI1} & = & -e^{a/2}\frac{a^{\prime}}{2}\delta_{iI}
  \,\,,\,\,
  i=I\geq 2\,,
  \label{spincon9}
\end{eqnarray}
and insertion into the Dirac equation (\ref{diraceq1}) leads to
\begin{eqnarray}
  & & 0=\Biggl(
  i\gamma^0e^{-b/2}\frac{1}{2}e^{b/2}\frac{b^{\prime}}{2}
  \gamma^0\gamma^1
  +i\gamma^1\partial_1
  \nonumber\\
  & & \quad
  +i\sum_{I=2}^d\gamma^Ie^{-a/2}
  \frac{1}{2}e^{a/2}\frac{a^{\prime}}{2}\gamma^1\gamma^I
  -m\Biggr)\psi
  \nonumber\\
  & & \quad
  =\left(i\gamma^1\left(\partial_1
  +\frac{b^{\prime}+(d-1)a^{\prime}}{4}\right)
  -m\right)\psi\,,
  \label{spincon10}
\end{eqnarray}
which is equivalent to Eq.~(\ref{diraceq2}).
\subsection{Energy-Momentum Tensor}
\label{EnergyMomentumTensor}
Similarly, the nonvanishing components of the energy-momentum tensor
(\ref{ergmonten1}) can be calculated as
\begin{eqnarray}
  T_{01} & = & \frac{ie^{b/2}}{4}\left(\bar\psi\gamma^0\partial_1\psi
  -(\partial_1\bar\psi)\gamma^0\psi\right)\,,
  \label{ergmonten4}\\
  T_{0i} & = &
  -\frac{ie^{(b+a)/2}}{8}(b^{\prime}-a^{\prime})
  \bar\psi\gamma^0\gamma^1\gamma^I\psi\delta_{Ii}
  \,\,,\, 
  i\geq 2\,, \, \, \, \, \, \,
  \label{ergmonten5}\\
   T_{11} & = & -\frac{i}{2}\left(\bar\psi\gamma^1\partial_1\psi
  -(\partial_1\bar\psi)\gamma^1\psi\right)\,,
  \label{ergmonten6}
\end{eqnarray}
Via the Dirac equation (\ref{diraceq2}), (\ref{diraceq3}) and the
Dirac algebra (\ref{diracalg1}) it now follows that the expression 
(\ref{ergmonten4}) identically vanishs, and the only
nonzero tensor components are indeed given by Eqs.~(\ref{ergmonten2}),
(\ref{ergmonten3}).
\subsection{The Constraint (\ref{diracsol12})}
\label{TheConstraint}
Although the constraint (\ref{diracsol12}) has to be fulfilled only by the
massless  Dirac field, we find it instructive to discuss this condition
also for the massive solutions (\ref{diracsol3}), (\ref{diracsol4}).
The general solution (\ref{diracsol6}) of the Dirac equation reads
in terms of the $2$-spinors entering Eq.~(\ref{diracsol12}) as
\begin{eqnarray}
	\chi_1 & = & \frac{1}{\sqrt{g}}
	\left(\left(
	\begin{array}{c}
		C_+\\
		D_+
	\end{array}
	\right)e^{-my}
	+\left(
	\begin{array}{c}
		C_- \\
		D_-
	\end{array}
	\right)e^{my}\right)
	\label{constraint1}\\
	\chi_2 & = & \frac{1}{\sqrt{g}}
	\left(\left(
	\begin{array}{c}
		D_+\\
		C_+
	\end{array}
	\right)e^{-my}
	-\left(
	\begin{array}{c}
		D_- \\
		C_-
	\end{array}
	\right)e^{my}\right)\,.
	\label{constraint2}
\end{eqnarray}
The massless case can be recovered by putting $m=0$ and replacing the function
$g$ according to Eq.~(\ref{diracsol10}).
It now straightforwardly follows
\begin{eqnarray}
	& & \chi^+_1\sigma^2\chi_1 =\frac{i}{g}\Bigl(
  \left(-C^*_+D_++D^*_+C_+\right)e^{2my}
	\nonumber\\
	& & \qquad\qquad\qquad\qquad
	+\left(-C^*_-D_-+D^*_-C_-\right)e^{-2my}
	\nonumber\\
	& & \quad
	-C^*_+D_-+D^*_+C_--C^*_-D_++D^*_-C_+\Bigr)\,,
	\label{constraint3}\\
	& & \chi^+_2\sigma^2\chi_2=\frac{i}{g}\Bigl(
	\left(-D^*_+C_++C^*_+D_+\right)e^{2my}
	\nonumber\\
	& & \qquad\qquad\qquad\qquad
	+\left(-D^*_-C_-+C^*_-D_-\right)e^{-2my}
	\nonumber\\
	& & \quad
	+D^*_+C_--C^*_+D_-+D^*_-C_+-C^*_-D_+\Bigr)\,,
	\label{constraint4}\\
	& & \chi^+_1\sigma^3\chi_1=\frac{1}{g}\Bigl(
	\left(C^*_+C_+-D^*_+D_+\right)e^{2my}
	\nonumber\\
	& & \qquad\qquad\qquad\qquad
	+\left(C^*_-C_--D^*_-D_-\right)e^{-2my}
	\nonumber\\
	& & \quad
	+C^*_+C_--D^*_+D_-+C^*_-C_+-D^*_-D_+\Bigr)\,,
	\label{constraint5}\\
	& & \chi^+_2\sigma^3\chi_2=\frac{1}{g}\Bigl(
	\left(D^*_+D_+-C^*_+C_+\right)e^{2my}
	\nonumber\\
	& & \qquad\qquad\qquad\qquad
	+\left(D^*_-D_--C^*_-C_-\right)e^{-2my}
	\nonumber\\
	& & \quad
	-D^*_+D_-+C^*_+C_- -D^*_-D_+ + C^*_-C_+\Bigr)\,,
	\label{constraint6}
\end{eqnarray}
showing that the conditions (\ref{diracsol12}) and (\ref{diracsol13}),
(\ref{diracsol14}) are indeed equivalent.

\end{document}